\documentclass[12pt,preprint]{aastex}

\shortauthors{Doeleman et al.}
\shorttitle{SiO Masers in Orion-KL}

\received{2003 October 31}
\begin{document}
\title{Using VLBI to Probe the Orion-KL Outflow on AU Scales}

\author{Sheperd S. Doeleman, Colin J. Lonsdale}
\affil{MIT Haystack Observatory}
\authoraddr{Off Route 40\\
       Westford, MA  01886;\\
       \em sdoeleman@haystack.edu, clonsdale@haystack.edu}
\author{Paul T. Kondratko}
\affil{Harvard-Smithsonian Center for Astrophysics}
\authoraddr{60 Garden St.\\
       Cambridge, MA 02138;\\
       \em pkondratko@cfa.harvard.edu}
\author{C. Read Predmore\footnote{Current address: Predmore Associates, 120
Pulpit Rd., Suite 22, Amherst, MA 01002}}
\affil{University of Massachusetts - Amherst}
\authoraddr{Dept. of Astronomy\\
       619 Lederle Graduate Rrch. Ctr.\\
       Amherst, MA 01003;\\
	\em predmore@PredmoreAssociates.com}

\slugcomment {Accepted for Publication in the Astrophysical Journal}

\begin{abstract}

We present the first contemporaneous 43GHz and 86GHz VLBI images of the v=1
J=2$\rightarrow$1 and J=1$\rightarrow$0 SiO masers in the Orion-KL nebula.
Both maser species exhibit the same general morphology of earlier
J=1$\rightarrow$0 maser images which appear to trace the edges of a bi-polar
conical outflow.  Surprisingly, the J=2$\rightarrow$1 masers form further from
the central protostar than the J=1$\rightarrow$0 masers, a fact not readily
explained by current SiO maser pumping models.  The average magnitude of
offsets between corresponding regions of the two masing transitions is
approximately 14\% of the total radial extent of the SiO maser emission.  This
offset indicates that each transition must trace different physical conditions.

\end{abstract}

\keywords {ISM: individual (Orion Kleinmann-Low)---ISM: jets and
outflows---masers---stars: formation---stars: winds, outflows}

\section {Introduction}

Young massive stars spend a considerable fraction (10-20\%) of their lifetimes
embedded within the molecular cloud cores from which they formed (Wood \&
Churchwell 1989).  Throughout this period, however, they can have a dominant
effect on the inter-stellar dynamics in the surrounding region by forming
complex and large scale molecular outflows.  The exact mechanism by which these
stars drive such outflows remains unclear, but models generally involve a
stellar wind entraining molecular material (Richer et al 2000).  Study of these
objects at early phases of evolution before they have emerged from parental
clouds is difficult due to the naturally large opacities in the IR and optical.
Furthermore, angular resolutions of instruments in these wavebands are
insufficient to image the very beginning of an outflow close to the young
stellar object.  Connected element radio interferometry relieves the opacity
problem, but still cannot probe angular scales much less than 0.5 arc seconds.
In some cases, high brightness and compact SiO maser emission is seen towards
massive star forming regions and can be imaged with 0.1 milli arc second
resolution (Greenhill et al 1998, Doeleman et al 1999 (Paper I), Eisner et al
2000).  The pumping requirements of these masers
($\rho_H\sim10^9\mbox{cm}^{-3}$, $T\sim1200K$ (Elitzur 1992)) require that they
originate very close to a high luminosity source and thus offer a method of
exploring the immediate circumstellar regions of select massive young stars
with angular resolutions much smaller than the stellar disk.  

The Orion BN/KL region is undeniably a site of intense molecular outflow
activity.  Bow shocks seen in $\mbox{H}_2$ form along ``fingers" extending up
to 2 arc minutes that collectively trace back to the center of the BN/KL region
(Gezari, Backman \& Werner 1998, Stolovy et al. 1998).  A high velocity and
weakly bipolar CO outflow is also centered there (Chernin \& Wright 1996).
Water maser features out to 15 arc seconds exhibit proper motions consistent
with a common center of expansion that is also near the BN/KL center (Genzel et
al 1979).  A number of objects in the region probably contribute to powering
this complex dynamical picture, but one in particular, the radio continuum
source I (Churchwell et al 1987), is associated with powerful SiO masers.  Its
inverted radio spectrum makes it likely that Source I marks the position of an
HII region or stellar jet associated with a young massive star, and precise
astrometry places I at the exact centroid of the Orion BN/KL SiO maser features
(Menten \& Reid 1995, Gezari et al 1998).  No definitive optical or IR
counterpart to Source I has been found, a fact attributable to the visible
extinction towards this object for which estimates yield values of
$A_{\nu}\sim60$ (Gezari et al 1998).

Compact v=1 J=1-0 SiO maser emission extends only $\sim70$AU from Source I,
well within the extent of the larger scale outflows described above.  Early
single dish polarimetry of this emission, combined with connected element array
observations, led to a model in which the SiO masers formed in a rotating and
expanding circumstellar disk (Barvainis 1984, Plambeck et al 1990).  These
efforts, however, were tainted by spectral blending.  Because the angular
extent of the maser emission is smaller than the synthesized beams of connected
arrays, widely separated maser features at similar radial velocities cannot be
distinguished and the maps of Plambeck et al (1990) show only the centroids of
emission in each observing frequency channel.  These maps show the centroids to
form two arcs that appear to encircle Source I, a morphology consistent with a
circumstellar disk.  

Recent high resolution VLBI imaging reveals that the SiO maser emission does
not form in two arcs, but resolves into four main regions that appear to trace
the outlines of a bipolar conical outflow oriented in the NW-SE direction of
the larger CO outflow (Greenhill et al 1998, Doeleman et al 1999).  The SiO
emission, however, is redshifted to the NW and blueshifted to the SE, opposite
the weak polarity of the CO outflow.  This apparent conflict can be accommodated
if the outflow is oriented close to our line of sight which would magnify the
effects of small direction changes in the outflow (Doeleman et al 1999).
Alternatively, limb brightening effects on the conical surfaces may allow blue
maser emission from a generally red shifted cone of emission (and vice-versa)
if the outflow alignment is close to the plane of the sky (Greenhill et al
1998).  In either case, it is important to investigate the physical conditions
and spatial extent of the maser emitting region in an effort to forge a link
between the small scale SiO structures and the larger scale outflows.  

In general, multi-transition SiO maser imaging holds the promise of revealing
small scale temperature and density gradients in the host environment.
Conclusions from previous multi-line SiO maser studies, though, have
necessarily been somewhat uncertain due to their heavy reliance on data from
single dish monitoring.  The particular difficulty with this approach stems
from the spectral blending that may occur when two spatially separated maser
features emit at nearly the same frequency.  Barvainis and Predmore (1985), for
example, used single dish polarimetry of the v=1 J=2$\rightarrow$1 and the
J=1$\rightarrow$0 transitions towards a set of evolved stars and inferred a low
degree of spatial overlap between the lines.  McIntosh and Predmore (1987),
however, used similar observations of SiO masers surrounding the variable star
Mira to conclude that the same transitions {\it were} cospatial.  Comparison of
single dish SiO maser spectra cannot definitively address the question of
relative maser positions.  Important connected-element interferometry work
exemplified by Baudry, Herpin \& Lucas (1999), Morita et al. (1992) and Colomer
et al. (1996) provides inter-line comparisons of centroids of emission at a
given velocity but no detailed spatial information.  The first multi-line VLBI
comparison of SiO masers was the claim by Miyoshi et al (1995) that the v=1 and
v=2 J=1$\rightarrow$0 transitions were largely cospatial towards the evolved
star VYCMa.  They concluded that only a collisional pumping mechanism could
account for the overlap. The angular resolution of their observations, however,
was still much larger than the smallest SiO maser feature sizes ($\sim 0.2$
mas), and thus their observations were insufficient to definitively prove
cospatiality.  Indeed, more recent and higher resolution observations (Desmurs
et al. 2000) suggest that on 0.2 mas scales, these two J=1$\rightarrow$0
transitions are offset from each other.  Phillips et al. (2003) have 
registered VLBI maps of the v=1 J=2$\rightarrow$1 and J=1$\rightarrow$0 maser 
emission in the envelope of the evolved star RCas.  For this source,  
some maser features from both transitions appear to arise in the same volumes
of gas, but the overall morphology of emission in the two transitions differs.   
 
Here, we report on results of mapping the Orion-KL SiO masers in both the v=1
$J=2\rightarrow1$ ($\nu_{\mbox{rest}}=86243.442$ MHz) and $J=1\rightarrow0$
($\nu_{\mbox{rest}}=43122.027$ MHz) transitions.  The accuracy of relative
astrometry in the images allows comparison at the sub-AU level and we find that,
at the resolution of our maps, the brightest masers in each transition are not
cospatial.  Spatial offsets between the two transitions place the
$J=1\rightarrow0$ emission slightly closer to the central exciting source,
leading us to conclude that they trace and occur in different physical
conditions.  

\section {Observations} 

We observed the Orion-KL SiO masers using the position determined by Wright et
al (1990) of $\alpha=5^h35^m14\fs505$, $\delta=-05^\circ22'30.45''$ (J2000).
Observations at $\lambda7$~mm took place on 13 Dec. 1997 using seven antennas
of the VLBA and one element of the VLA, both run by the NRAO\footnote{The
National Radio Astronomy Observatory is a facility of the National Science
Foundation operated under cooperative agreement by Associated Universities,
Inc.}.  Paper I describes the $\lambda$7~mm observations and the resulting
calibration and imaging steps used to generate high resolution maps.
Coordinated Millimeter VLBI Array\footnote{Support for the Coordinated
Millimeter VLBI Array work at the Haystack Observatory is provided under a
grant from the NSF to the Northeast Radio Observatory Corporation} (CMVA)
observations at 86~GHz covered a time range from 13 Dec. to 15 Dec. 1997 and
included the following antennas: Haystack(Westford, MA.), FCRAO 14m (Amherst,
MA.)\footnote{This work was supported by NSF grant AST 97-25951 to the Five
College Radio Astronomy Observatory}, Kittpeak 12m(Kittpeak, AZ.), the phased
BIMA (Redding, CA.) and the VLBA site at PieTown, NM.  The overlap in time
between the 43GHz and 86GHz observing sessions ensures that variability of
source structure has a negligible effect on comparisons between the two maser
transitions.  The array was split into two sub-arrays, a 'low-resolution' array
consisting of the relatively short (85~km) Haystack- Quabbin baseline and a
'high-resolution' array comprising Kittpeak, BIMA and PieTown.  There were very
few interferometric detections between the short baseline array and the high
resolution array, forcing us to separate the analysis of the two data sets.  A
technical problem at the PieTown VLBA site rendered data from the
'high-resolution' array unsuitable for imaging.  In this letter, we report on
results only from the Haystack-Quabbin baseline.  Data were recorded in three
partially overlapping IF channels of the MKIII VLBI system, each of 4MHz
bandwidth yielding a total effective velocity coverage at 86~GHz of
$\sim34$km/s.  The IF overlap was sufficient to avoid band edge effects and
also allowed removal of instrumental phase shifts between IFs.  Correlation at
Haystack Observatory yielded 112 spectral channels in each IF for a velocity
resolution of 0.12km/s which closely matches the resolution of the 43~GHz
results (0.11km/s) of Paper I.

\section {Calibration and Imaging}

Single baseline 86~GHz VLBI has already been used to map SiO masers around the
evolved giant star VXSgr (Doeleman, Lonsdale \& Greenhill 1998) and the
associated difficulties are well understood.  The most important aspect of the
reduction is the need to find a spectrally isolated component which is
point-like.  Phase referencing the entire data set to this reference feature
simultaneously removes atmospheric effects and shifts the reference feature to
the map origin.  Without a point source as reference, the data set would be
corrupted by the emission structure in the reference channel which cannot be
satisfactorily determined in the absence of calibrated phases.  To identify a
point source with a single baseline, one must search through all spectral
channels for one in which the visibility amplitudes are constant as a function
of baseline length and orientation.  This signature can easily be masked by
amplitude variations due to antenna gain fluctuations and high quality antenna
gain calibration is therefore essential.

Gain calibration for both sites was obtained using spectral template fitting.
A 90 second total power spectrum of the maser emission at the Quabbin telescope
was calibrated assuming a 40 Jy/K gain and served as a template.  Spectra at
all other times from both sites were fitted to the template, and relative gains
as a function of time for both antennas were determined
(Fig.~\ref{fig:gaincurves}) for each 90 second interval.  Relative calibration
errors are at the 5\% level due to the high quality of the total power spectra
while we estimate the absolute calibration of the template spectrum to be
within 10\%.  Once calibrated, the channel at $V_{\mbox{\footnotesize
LSR}}$=0.84km/s had amplitudes constant to 20\% and we adopted it as the
reference.  For comparison, Fig.~\ref{fig:amplitudes} shows the amplitudes as a
function of time for the reference channel as well as amplitudes at a nearby
velocity which exhibit dramatic ``beating" indicative of complex structure.

Strong fringe detections on a single 390 second VLBI scan of the continuum
source 3C273 provided the delay calibration for both days allowing removal of
linear phase slopes as a function of frequency across the bandpass.  Phase
offsets between adjacent IFs were checked using maser features in the
overlapping portions of the three IF channels and shown to be good to within a
few degrees.  In addition, visibility phase versus frequency was differenced for
two segments of data at the same LST on both days and showed a negligible shift
in delay.  The lack of phase slope in this difference validates use of the
single 3C273 scan to calibrate the delay for both days.  

Fringe rate solutions from the reference channel were applied to the data and
synthesis image cubes constructed using standard techniques implemented in the
NRAO AIPS software package.  Resulting images covered
$0.64\arcsec\times0.64\arcsec$ on the sky, sufficient to completely map the
known extent of the maser emission (Paper I).  Due to high beam sidelobe levels
(71\%), CLEAN deconvolution was applied conservatively in each velocity channel
using a loop gain of 0.05 with a limit of 100 iterations.  Deconvolution tests
on these data showed that the adopted CLEAN parameters prevented divergence in
the algorithm and minimized imaging artifacts.  The restoring beam was $9\times
66$ milli arcseconds with a PA of $-18^\circ$.  

\section {Image Analysis}

With only two antennas and the relatively sparse baseline coverage that
results, the fidelity of the channel maps is limited by sidelobes of bright
emission regions caused by imperfect deconvolution of the synthesized beam.
Bright isolated maser emission, for example, is often accompanied by negative
sidelobes which distort nearby faint emission.  Faint emission that is far
removed from bright map features can be more clearly distinguished against the
residual map background.  Setting a threshold above which map features are taken
to represent maser emission, is thus position dependent and requires that each 
channel map be considered separately.

Composite maps for three velocity ranges corresponding to the three observed
4MHz passbands were formed by selecting the maximum intensity of all velocity
channels at each image pixel (Fig.~\ref{fig:overlay}a-c).  The lowest 3mm
contours in the composite maps mark the cutoff below which sidelobe features
begin to appear. 

For a more detailed analysis of the 3mm image, maser spot maps were made at
each velocity, where 'spot' (or component) will be defined as an isolated
region of maser emission within a single frequency channel.  This was done by
fitting 2-D elliptical Gaussians to each map feature to determine size and
integrated flux density.  In almost all cases, the high spectral resolution and
relatively coarse angular resolution of the observations ensured that each
identified map feature could be well modeled as an unresolved source convolved
with the restoring beam.  Restrictive criteria were applied to these features
to identify maser emission components.  These included feature persistence in
at least two adjacent velocity channels and a flux density threshold set by the
largest negative sidelobe within 3 beamwidths.  In Fig.~\ref{fig:overlay}d, all
3~mm maser spots meeting our selection criteria are plotted as circles with
area proportional to total flux density.   

The positional accuracy of maser components relative to each other within the 3mm
map depends on a number of factors.  Foremost among these is the effect of
non-point like structure in the reference channel which will contaminate phases
for all other channels resulting in position errors.  In the Orion 3mm data
(Fig.~\ref{fig:amplitudes}) the observed variation in reference channel
visibility amplitudes is $\sim$20\%.  The corresponding phase error can be
estimated by assuming this amplitude variation is due to two spatially
separated maser spots in the reference channel.  The observed range of
amplitudes would then imply visibility phase variations up to $\pm6^\circ$ or
$\pm0.02$ of the synthesized beam width.

A secondary concern is the effect of combining the two days of 3mm VLBI data on
relative maser positions.  The data were combined primarily to increase signal
to noise ratio, but if the masers exhibit very high proper motions, data
between the two days could be inconsistent.  Assuming the proper motions are
comparable to the radial velocity range observed ($\sim 30$km/s), the two day
separation corresponds to motions of 0.1 mas, a negligible offset compared to
the 3mm VLBI resolution.  Even if the SiO outflow were $\sim10^\circ$ out of
the plane of the sky, the corresponding proper motions would be $\sim150$ km/s, 
leading to 0.5 mas positional offsets between the two days.

Mis-calibration of the interferometer delay due to geometrical and station
clock errors can also lead to uncertainty in relative astrometry (Genzel et al
1981, Thompson, Moran \& Swenson 1986).  Such delay errors cause phase slopes
across the bandpass leading to astrometry errors whose magnitudes vary as a
function of frequency.  The high signal to noise ratio 3C273 detections limit
clock errors to $\sim5$ns or 18 degrees of phase across the observing bandpass.
The maximum geometrical delay error, in turn, is set by the uncertainty in the
position of the Orion BN/KL SiO masers and the baseline length.  The Orion
BN/KL maser position is known to within $0.6\arcsec$ ($3\sigma$) which, coupled
with the relatively short Quabbin - Haystack baseline, produces a delay error
of only $0.75$ns.  Thus, the total delay error contributes positional errors of
no more than 21 degrees of phase, or $\pm 0.06$ synthesized beam widths. 

Further errors in maser spot positions (Fig.~\ref{fig:overlay}d) can be
attributed to uncertainty in identification of individual features and fitting
them with elliptical Gaussians, but in all cases, these uncertainties are a
small fraction of the synthesized beam.  Combining all sources of relative
positional error above, we conservatively limit relative errors in maser spot
positions of 0.1 of the synthesized beam.  For the main purposes of the present
work, the combination of all relative positional errors are small compared to
the errors in aligning the 3mm and 7mm maser transitions discussed below.

\section {Comparison of 3mm and 7mm Emission}

\subsection{Characterization of 3mm Maser Emission}

Maps of the 3mm maser emission show it to be broadly similar in structure to
that observed in the 7mm transition.  Regions A, B, G and H marking the
outlines of opposing conical outflows at 7mm in Paper I have corresponding
regions at 3mm.  Even region E, which does not conform to the simple bi-conical
picture, is populated by both SiO transitions.  At 3mm, each region comprises
multiple components whose velocity widths range from 0.5 km/s to over 2 km/s.
These are larger than spectral widths of typical maser features in the 7mm
image, but this is likely due to the larger beam size at 86~GHz which subtends
multiple closely spaced features at similar velocities.  As at 7mm, there is
obvious velocity overlap between region pairs (A,B) and (G,H); strong emission
often appears in both paired regions at the same velocity.  

\subsection{3mm and 7mm Map Alignment}

Because the 3mm and 7mm data were obtained using different instruments, it is
impossible to register the maps using phase referencing techniques.  We
note that use of these techniques may be possible in the future as experience with
VLBA operation at 86~GHz increases and if the frequency switching time remains
below the coherence time of the atmosphere.  Instead, we necessarily explored
registration techniques involving map comparisons and certain assumptions about
maser structure.  

Comparison of the two maps reveals that a pure translation of one map relative
to the other will not allow all four main emission regions (A, B, H, (F+G)) to
coincide.  Exactly superposing region B, for example, at each frequency leaves
the remaining regions grossly misaligned.  If, however, we assume that emission
in both transitions originates in similar interactions of a bi-polar outflow
with the surrounding cloud, then the center of symmetry in both transitions
should be coincident and reflect the position of the protostar.  An approximate
center of symmetry in each map can be defined as the intersection of lines that
connect the emission centroids of regions A with H and B with (F+G).  
Registering the 3mm and 7mm emission in this manner (Fig.~\ref{fig:overlay}), shows that
J=2$\rightarrow$1 masers appear to form farther from the central proto-star
than the J=1$\rightarrow$0 masers.  

After registration, total offsets between the 3~mm and 7~mm centroids for each
emission region were measured to be: A (6.6 AU), B (7.5 AU), F+G (7 AU), H (17
AU) assuming a 450 pc distance to the source (Paper I).  Variation in the
offsets is directly related to the extent of emission in each region.  Region
H, for example, exhibits the largest offset, due primarily to the large area
over which 3mm emission is found.  In all cases, the 3~mm emission appears
further from the presumed location of the protostar than the 7~mm emission.

\subsection {Spectra}

Total power spectra of each transition in Fig.~\ref{fig:total_spectra} show
that the 3mm and 7mm masers share the same basic double peaked form with a
central velocity near 5.3 km/s.  This indicates a common bulk flow affecting both
maser lines.  In fact, individual spectra of emission in regions A-H of the 3mm
map cover roughly the same velocity ranges as the corresponding 7mm spectra :
Region A (11 $\rightarrow$ 17.5 km/s), Region B (13.5 $\rightarrow$ 19 km/s),
Region H (2.5 $\rightarrow$ -9 km/s), Region G ( -2.5 $\rightarrow$ -11 km/s).
Both the red and blue shifted peaks contain multiple spectral components, a few
of which have matching peaks in both transitions that correspond to within 0.5
km/s.  A one-to-one correspondence in spectral features between the two
transitions, though, is clearly absent.  The observed spatial offset between
the lines underlines the pitfalls of deriving spatial coincidence information
solely from single dish spectra.  

\section {Discussion}

\subsection {Outflow Model and Dynamics}

Impetus for the biconical outflow model of the Orion-KL SiO masers originated
with a clear "X" morphology of the v=1 J=1$\rightarrow$0 transition, presumed
to trace the outflow boundary.  Though generally not coincident with the
J=1$\rightarrow$0 masers, the J=2$\rightarrow$1 emission displays a remarkably
similar general structure consistent with the outflow model.  This general
structure is consistent with both SiO maser transitions inhabiting a zone of
shocks and overdensities where an outflow interacts with the surrounding
molecular medium (Paper I).  The resolution of the J=2$\rightarrow$1 image does
not allow us to compare the two transitions on the smallest scales seen in the
J=1$\rightarrow$0 maps.  We cannot rule out some overlap on scales smaller than
the 3~mm beam and future high resolution work at this frequency is needed to
address this.  One caveat to make clear is that in both transitions, roughly
half the flux density is undetected by VLBI.  This ``missing flux" must exist
in structures larger than the 5 milliarcsecond synthesized beam of the
J=2$\rightarrow$1 data.  The smallest baselines in the VLBA array also
correspond to this size scale.  Detection and location of this large scale
emission will require connected element arrays with baselines of order 100 km.  

The combined spatial and velocity information from both 3~mm and 7~mm VLBI now
make it possible to explore velocity patterns within each main emission region.
In region A, the combined 3~mm and 7~mm maser emission covers a range of radii
from 40 to 67 AU as measured from the central registration point.  Over this
range, the average radial velocity across region A steadily decreases from 26
to 13 km/s implying a smooth velocity gradient of $\sim 0.5$km/s/AU due North.
Similar calculations for the other three regions yield no clear velocity
gradients, arguing for a local explanation of the pattern observed in Region A.  

One possibility is that in Region A, maser features closer to the protostar are
also located closer to the central stellar jet.  Molecular gas closer to the
jet will accelerate more quickly and to generally higher terminal velocities,
giving rise to the observed velocity gradient.  In this scenario, the thickness
of the conical shell where masers occur is sufficient to generate the velocity
difference among Region A maser features.  Similarly, a local gradient in the
density of the medium surrounding the stellar jet could have the same effect on
the radial maser velocities by altering the efficiency of entrainment as a
function of distance from the central protostar.  Constraining these
possibilities is not possible with the measurements described herein, but
future proper motion measurements of individual maser features will allow a
much more detailed examination of the SiO maser dynamics and directly address
this issue.  For completeness, we note that one could also attribute the
velocity gradient to a gravitational deceleration of a ballistic flow in which
the masers form.  The central mass required would then be $M_\star \sim
28\;[\cos^2(\theta_{\mbox{los}})\sin(\theta_{\mbox{los}})]^{-1}\; M_\odot$
where $M_\star$ is the enclosed central mass and $\theta_{\mbox{los}}$ is the
angle made by the direction of propagation of the Region A SiO masers to our
line of sight.  At its minimum, this expression yields a central mass of
$M_\star \sim 73 M_\odot$.  In the case of gravitational deceleration, though,
one would expect similar velocity gradients in all maser regions.  

The new J=2$\rightarrow$1 images also highlight a region not directly addressed
in Paper I.  Region E is the weakest of all the labeled maser complexes but
appears clearly in J=2$\rightarrow$1 and more faintly in J=1$\rightarrow$0.
This region differs importantly from the others in that the spatial
relationship between the transitions is reversed with J=1$\rightarrow$0
slightly farther from the center.  The mere existence of this maser cluster in
both transitions is difficult to understand in the context of a pure biconical
outflow model.  It cannot correspond in any simple way to a conical edge or
tangent line.  Since Region E emission velocities are near the centerpoint
between the red and blue lobes (Vlsr$\sim$5 km/s) it may result from outflow
shocks propagating to a central disk or torus oriented perpendicularly to the
outflow as observed by Wright et al (1995) in thermal SiO.  Another possibility
is that Region E forms in an equatorial outflow modeled by Greenhill et al.
(1998) to explain the morphology of the $\mbox{H}_2\mbox{O}$ masers.  Whatever
its genesis, the 86~GHz confirmation of strong SiO emission in Region E must be
accounted for in future dynamical models.

\subsection {Maser Pumping}

Most treatments of SiO maser pumping model the case of a spherically symmetric
evolved star.  Elitzur (1982) has outlined a possible explanation for the
special case of the Orion-BN/KL SiO masers assuming they form in the bulk of an
expanding wind from the protostar.  Models of this type are not generally
workable in the evolved stellar case as mass loss rates in those objects are
too small to sustain maser amplification within a smooth wind.  The validity of
the Elitzur (1982) model, which also dealt with the specific radiation field in
the region, is now called into question by the filamentary maser features seen
in J=1$\rightarrow$0 (Paper I).  These high aspect ratio structures indicate
maser formation in shocks and local density enhancements rather than in a
smooth wind.

Many stellar SiO maser models incorporate the effects of dust grain formation
and stellar pulsation driven shocks, effects that likely play a role in the
Orion SiO masers.  Both radiative and collisional pump mechanisms have been
used to explain SiO masers in these models, but a general consensus remains
elusive.  VLBI results of Miyoshi et al (1994) showing the J=1$\rightarrow$0,
v=1 and v=2 masers in VYCMa and WHya to be coincident within 2 mas,
strengthened the case for collisional pumping which allows these transitions to
be cospatial over a wide range of physical parameters (Lockett \& Elitzur
1992).  Radiative pumping schemes are much more restrictive and one would not
typically expect such coincidence (Bujurrabal 1994).  Desmurs et al. (2000)
have mapped TXCam with a resolution of 0.2 mas and claim that centroids of
J=1$\rightarrow$0, v=1 and v=2 maser features are offset by $\sim$1.5 mas.
Their refutation of the Miyoshi et al. (1994) results must be tempered though,
by the fact that their maps show a high degree of overlap between the
transitions despite the measured mean offset.

The positional offset now observed between the J=2$\rightarrow$1 and
J=1$\rightarrow$0 masers is in stark disagreement with predictions of most
maser pumping models for SiO masers around evolved stars.  Radiative and
collisional excitation theories both predict multiple SiO maser lines among
rotational levels within a given vibrational state.  These rotational ``chains"
are due to the monotonic decrease in radiative decay rates with increasing
rotational level (J) when ro-vibrational transitions become optically thick
(Elitzur 1992).  This produces a natural inversion between rotational states
given a J-independent pump mechanism.  In such a case, one would expect v=1,
J=2$\rightarrow$1 and J=1$\rightarrow$0 masers to generally inhabit the same
volumes of gas.  Our results show that this is not the case for the brightest
maser features at the resolution of our maps.  

Some SiO maser models do allow for spatial separation of rotational masers
within a vibrational state, but require specialized conditions for this to take
place.  Lockett \& Elitzur (1992) show that collisional pumps tuned for
conditions around evolved stars selectively quench higher J level transitions
as SiO column densities rise, so that the J=1$\rightarrow$0 maser is the only
one to survive above $10^{20}\mbox{cm}^{-2}$.  These results are consistent
with J=1$\rightarrow$0 masers occuring closer to the protostar where higher
neutral $H_2$ densities would be found.  Larger SiO column densities might also
result from an increase in SiO abundance due to the liberation of SiO into a
gaseous state from dust grains in shocks at the interface between the Orion
outflow and the surrounding medium (Caselli, Hartquist \& Havnes 1997).  Such
interpretations would, however, require very high column densities to suppress
the higher J transition.  Radiative pumps typically show little change in
J=1$\rightarrow$0 and J=2$\rightarrow$1 maser intensities as $H_2$ density and
SiO abundance are varied (Bujurabbal 1994) and offer no clear explanation of
the findings reported here.    

Given the large number of SiO maser transitions possible, saturation and
competitive gain effects can be important.  Doel et al (1995) argue that SiO
maser spot sizes and flux densities imply that much of the maser emission is in
the saturated regime.  These authors have extended SiO maser models for evolved
stars to include these effects and find that the J=2$\rightarrow$1 maser gains
can exceed those of J=1$\rightarrow$0 in a very narrow range of $H_2$ density
around $5\times10^9\mbox{cm}^{-3}$.  The possibility that this density range
holds throughout the large SiO maser region in Orion BN/KL is unlikely.

Potentially of more relevance to the Orion BN/KL case is the work of Humphreys
et al (2002) to include the effects of hydrodynamic shocks on SiO maser
formation in evolved stellar envelopes.  In this model, stellar pulsations
drive shocks which enhance collisions and give rise to pockets of SiO masers
emission.  Simulations show that the v=1 J=2$\rightarrow$1 emission occurs at
slightly larger radii from the central star than the v=1 1$\rightarrow$0
emission.  The size of this effect ($\sim2$\% of the maser radii) is much
smaller than the $\sim14$\% offset between the two transitions we have observed
here.   Inclusion of shocks in SiO models, though, will likely be important 
in future specific application to the Orion BN/KL case.

In general, the relevance of these SiO maser models in the context of hard
radiation fields and shocks in the environment surrounding Source I is
questionable.  Perhaps the clearest statement that can be made regarding the
offset between the two rotational maser transitions discussed in this work, is
that they appear to require distinct physical conditions for maser
amplification.  Use of these transitions as probes of the Orion-KL environment
depends heavily on the specific predictions of theoretical models which
currently cannot adequately explain our observations.

\section{Conclusions}

We have made the first contemporaneous spectral line VLBI observations of 3mm
and 7mm wavelength SiO maser transitions towards the Orion BN/KL region.
Images of the 3mm v=1 J=2$\rightarrow$1 transition show the masers to be
grouped in four main emission regions along the arms of an 'X', similar
morphology to that previously reported for corresponding observations of the
7mm J=1$\rightarrow$0 transition.  These results reinforce a scenario in which
the SiO masers form in the interface region between a bi-conical protostellar
outflow and the surrounding medium.  Long tangential maser gain paths along the
edges of the outflow result in the masers appearing along the outline of the
outflow.  Significant SiO maser emission outside the outflow cones defined by
the main maser regions is easily identified in the new J=2$\rightarrow$1 image
(Region E) and exists, but is less distinct, in the J=1$\rightarrow$0 map.
These maser features are inconsistent with the simple bi-conical picture and
may indicate the presence of a protostellar disk or other dense outflowing
material in the plane orthogonal to the stellar outflow.

In contrast with predictions of SiO maser models, we observe a positional
offset between the centroids of J=2$\rightarrow$1 and J=1$\rightarrow$0 maser
emission with the higher rotational transition occurring farther from the
central protostar.  This offset indicates a preference of the two transitions
for distinct physical environments.  The J=2$\rightarrow$1 masers extend the
maximum radius at which SiO masers are seen in Orion-BN/KL to 67 AU and agree
with a velocity gradient observed in J=1$\rightarrow$0 emission along the
Northern outflow limbs.  This is almost certainly due to an effect local
to this region of emission as the remaining three maser regions exhibit no
clear velocity gradients.

A more complete picture of the SiO masers and their use as tracers at the
origins of the Orion BN/KL outflow will require further observations.  Proper
motion studies of individual maser features will produce three dimensional
velocity information and allow detailed study of dynamics in the maser region.
High resolution (0.2 mas) simultaneous observation of multiple maser lines are
needed to make progress toward understanding the maser pump models.  Indeed,
VLBI study of SiO masers has now outpaced theoretical maser efforts.  The utility
of multi-line studies highlights the need to extend the spectral line VLBI
technique to higher frequencies to reach other SiO maser transitions.  Imaging
of higher rotational level maser lines may reveal unexpected effects similar to
those discussed in this work.

\clearpage

\section{Figure Captions}

\figcaption{Gaincurves in units of System Equivalent Flux Density determined for
the Haystack and Quabbin antennas using a total power spectrum template method
averaged over 360 second intervals.  Calibration points from day 1 are marked by
filled circles, those from day 2 by open circles.  Antenna gain is determined on the
target maser source so all pointing errors and opacity effects are corrected
for.  In the actual data reduction, antenna gains were calculated for every 90
second interval.  \label{fig:gaincurves}}

\figcaption{Calibrated visibility amplitudes at two different velocities.  Day
1 marked with filled circles, day 2 by open circles.  The lower panel shows
variation in the amplitude indicative of complex structure.  The upper panel
shows the amplitudes for the channel chosen as the phase reference.  The near
constant reference amplitudes are consistent with a point like brightness
distribution.  The degree to which structure in the reference channel deviates
from a point source will add to positional errors in the final map.
\label{fig:amplitudes}}

\figcaption{Comparison of v=1 J=2$\rightarrow$1 and
J=1$\rightarrow$0 masers in Orion-KL.  Panels a through c show
overlayed contour maps of both transitions (J=2$\rightarrow$1 in green
and J=1$\rightarrow$0 in black) summed over different velocity
ranges.  All contour levels are spaced by powers of $\sqrt{2}$.  Contour ranges
for J=2$\rightarrow$1 are: a) 34 to 187 Jy/Beam; b) 12 to 92 Jy/Beam; c) 45 to
180 Jy/Beam.  Contours for J=1$\rightarrow$0 transition maps are: a) 42 to 469
Jy/Beam; b) 28 to 315 Jy/Beam; c) 57 to 634 Jy/Beam.  The J=1$\rightarrow$0
emission has been convolved with the J=2$\rightarrow$1 beam for ease of
comparison.  Panel d shows the J=1$\rightarrow$0 emission summed over all
velocities compared with J=2$\rightarrow$1 maser spots obtained by fitting
elliptical Gaussians to map features.  Circle areas are proportional to total
flux density, and circle color represents velocity.  A central red ``X" marks
the map registration point (and presumed protostar location) determined by the
method described in the text.  Arms of the ``X" are oriented to show the
outflow opening angles determined by lining up centroids of the J=2$\rightarrow$1
emission with the map center.  Regions labeled A,B,E,F,G and H correspond
to similar regions defined in Paper I.  \label{fig:overlay} }

\figcaption{Simultaneous spectra of the v=1 J=2$\rightarrow$1 and
J=1$\rightarrow$0 SiO maser transitions towards Orion-KL.  The similar
double-peaked forms suggest that both transitions are subject to the same large
scale outflow velocity pattern.  Spectral dissimilarities are probably due to a
spatial offset between the two transitions as determined with long baseline
interferometry.  \label{fig:total_spectra} }

\begin{figure}[p]
\epsscale{0.60}
\plotone{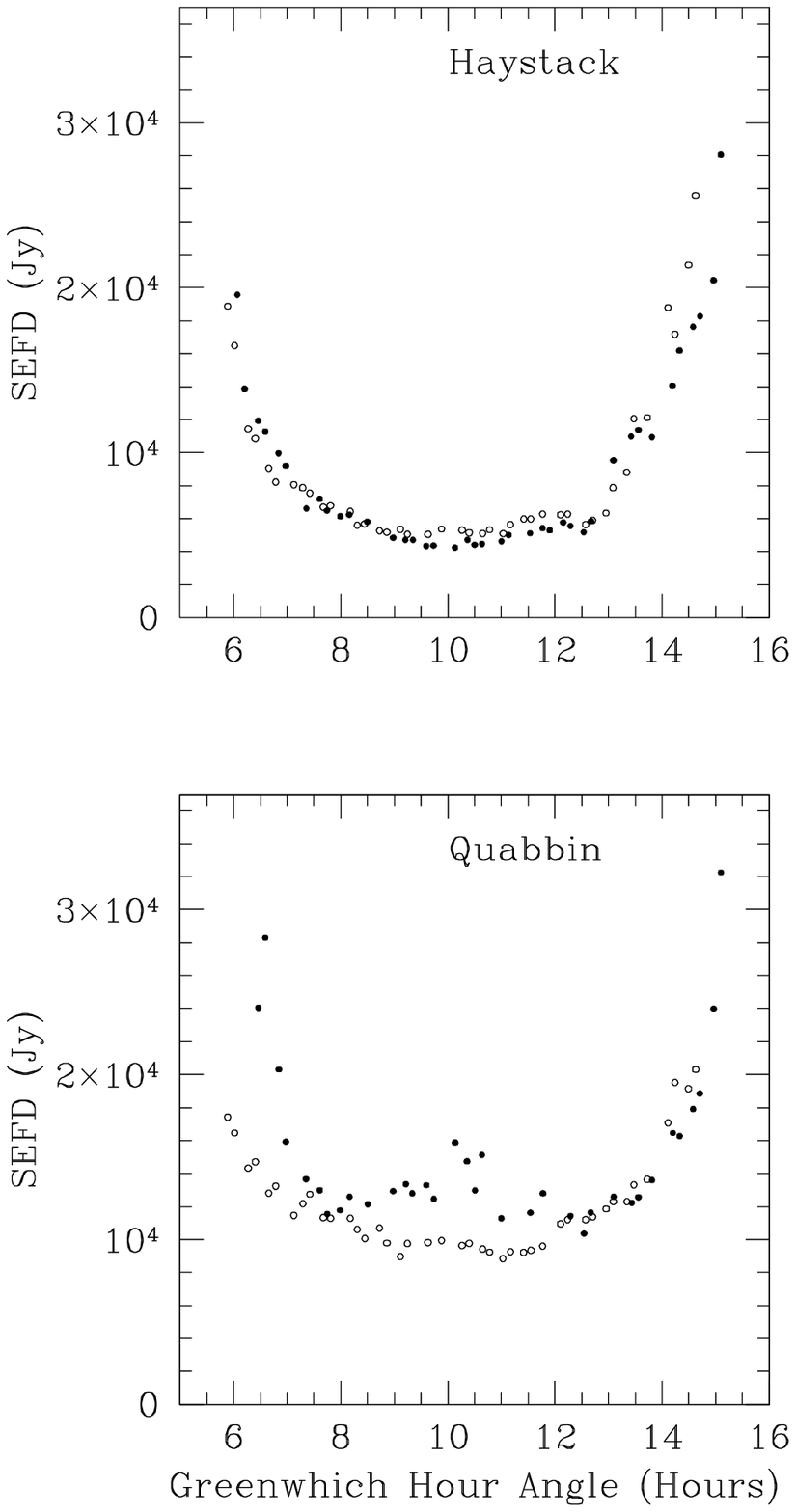}
\end{figure}

\begin{figure}[p]
\plotone{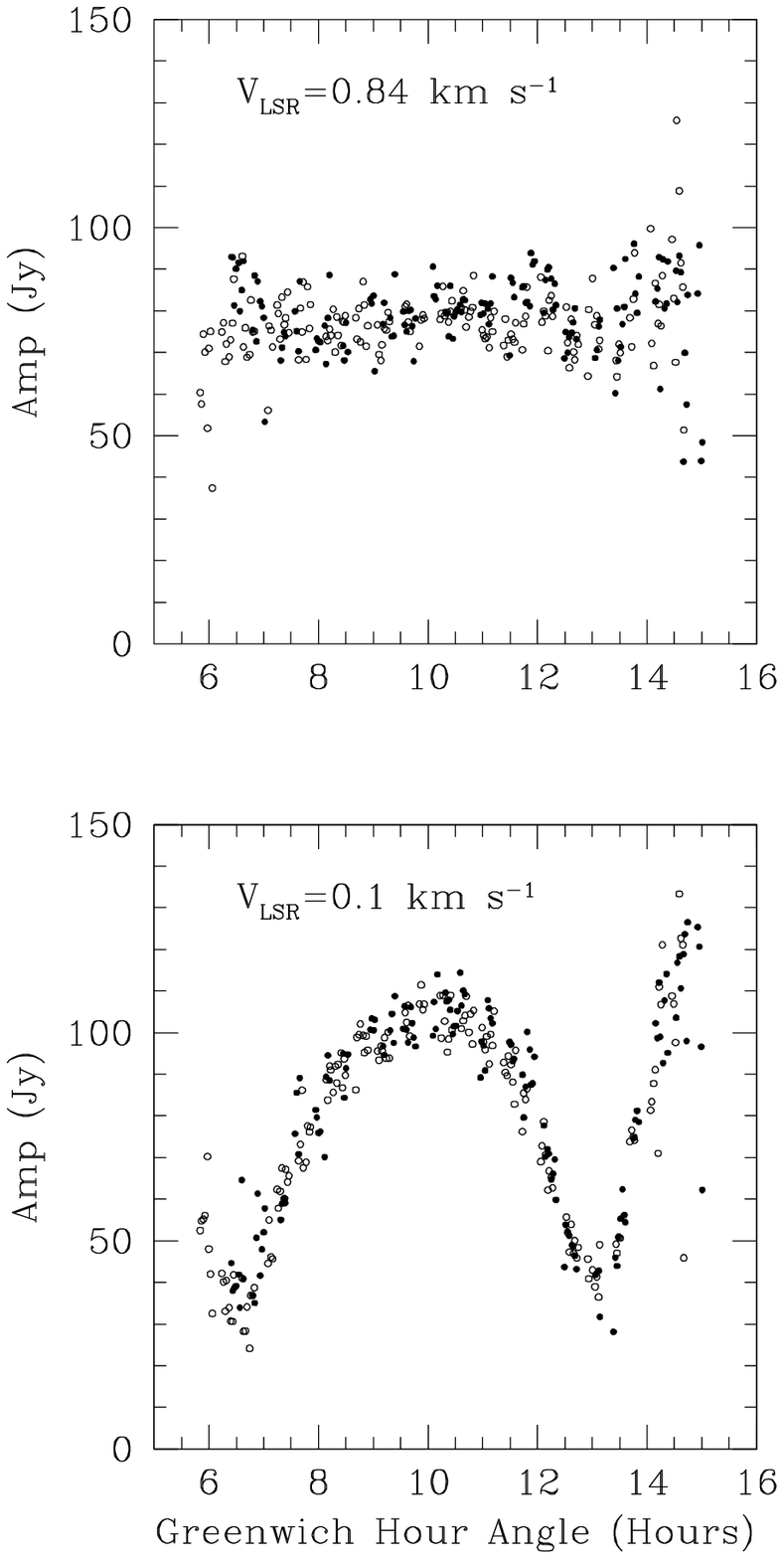}
\end{figure}

\begin{figure}[p]
\epsscale{1.00}
\plotone{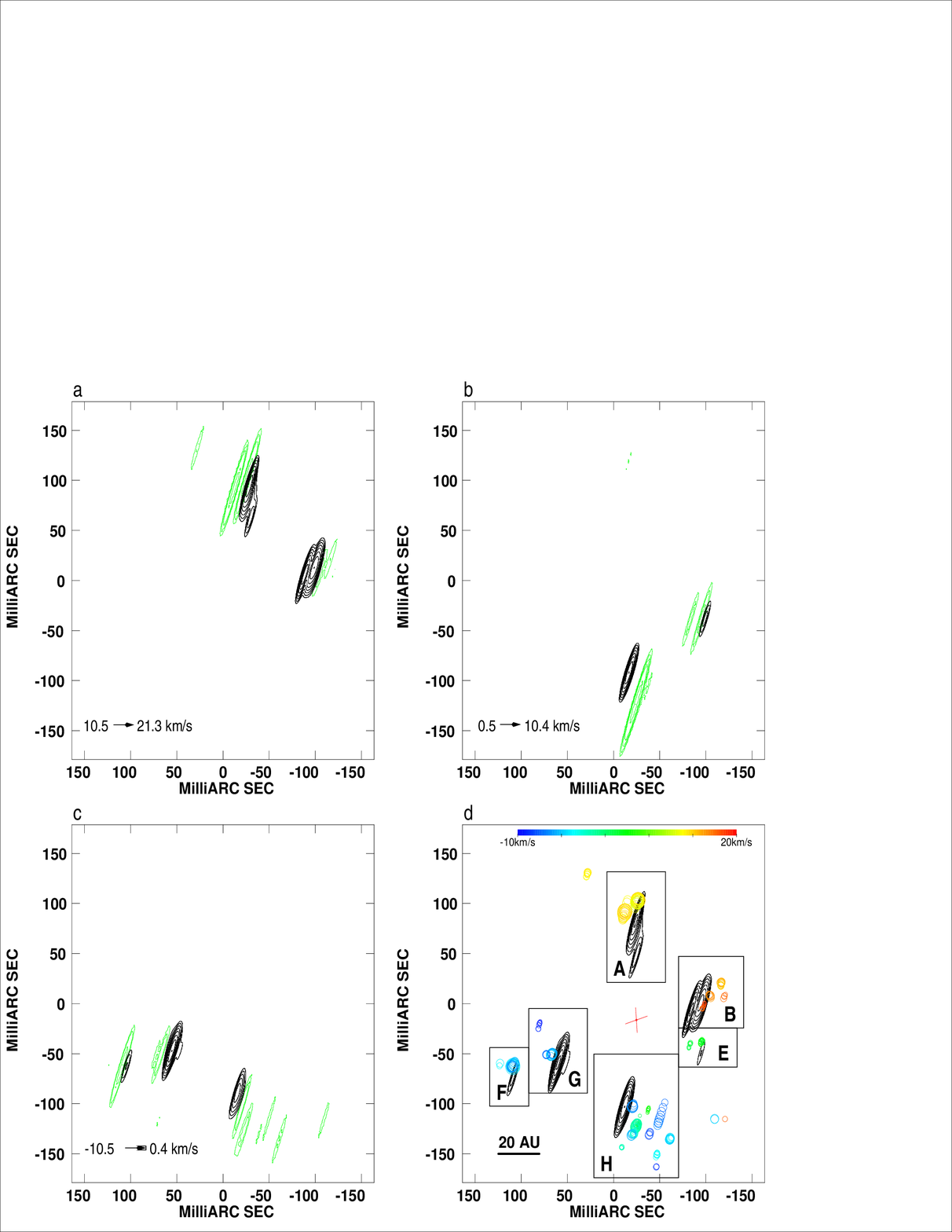}
\end{figure}

\begin{figure}[p]
\epsscale{1.00}
\plotone{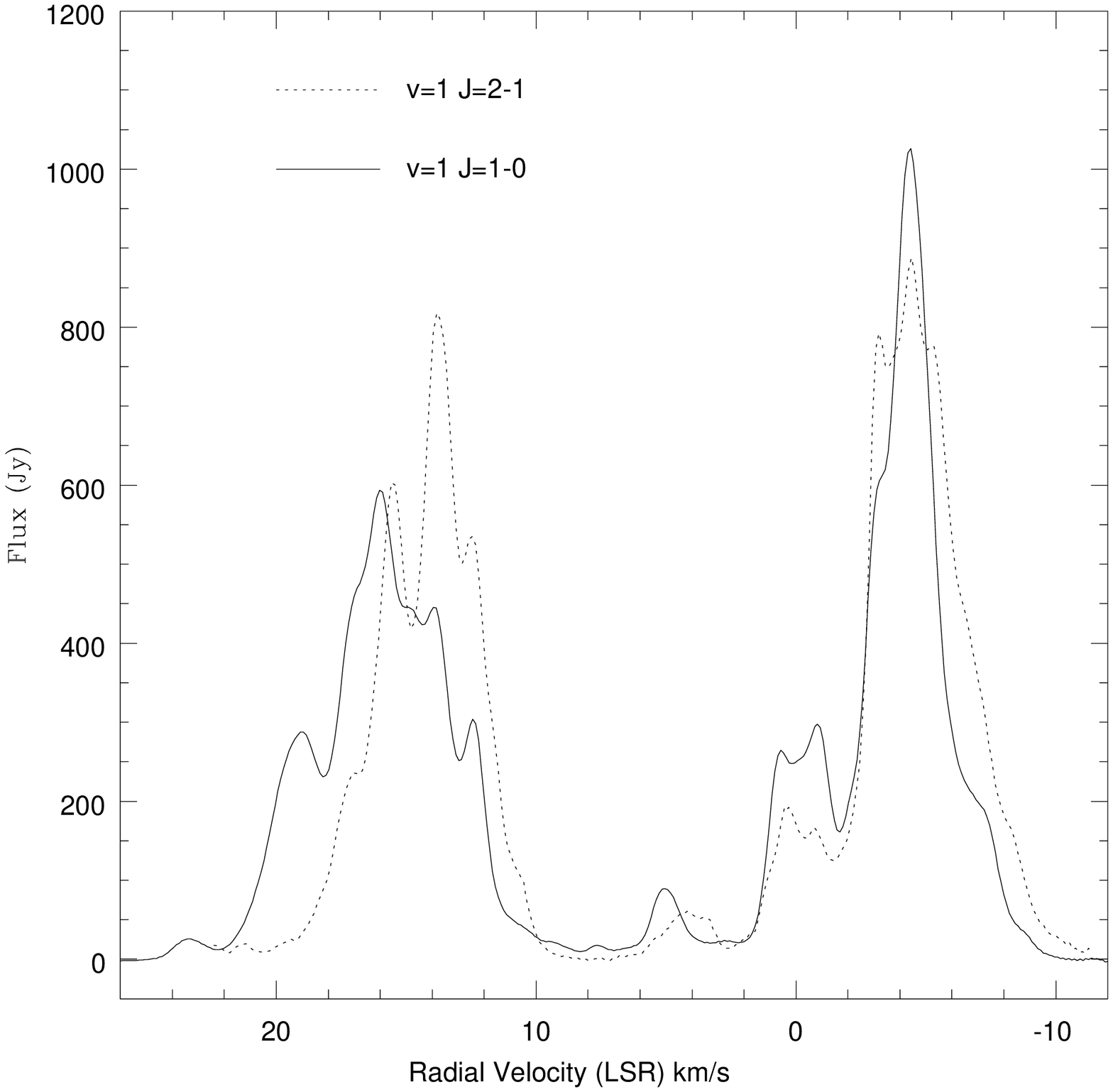}
\end{figure}

\end {document}